# Domain wall-grain boundary interactions in polycrystalline Pb(Zr$_{0.7}$Ti$_{0.3}$)O$_3$ piezoceramics

J. Schultheiß[1,2], S. Checchia[3,4], H. Uršič[5], T. Frömling[1], J. E. Daniels[6], B. Malič[5], T. Rojac[5], and J. Koruza[1,*]

[1] Department of Materials and Earth Sciences, Technical University of Darmstadt, Alarich-Weiss-Straße 2, 64287 Darmstadt, Germany

[2] Department of Materials Science and Engineering, Norwegian University of Science and Technology (NTNU), 7034 Trondheim, Norway

[3] European Synchrotron Radiation Facility (ESRF), 71 Avenue des Martyrs, 38000 Grenoble, France

[4] MAX IV Laboratory, Lund University, Box 117, 221 00 Lund, Sweden

[5] Electronic Ceramics Department, Jožef Stefan Institute, Jamova cesta 39, 1000 Ljubljana, Slovenia

[6] School of Materials Science and Engineering, University of New South Wales, 2052 Sydney, Australia

[*] Corresponding author: koruza@ceramics.tu-darmstadt.de

**Abstract**

Interactions between grain boundaries and domain walls were extensively studied in ferroelectric films and bicrystals. This knowledge, however, has not been transferred to polycrystalline ceramics, in which the grain size represents a powerful tool to tailor the dielectric and electromechanical response. Here, we relate changes in dielectric and electromechanical properties of a bulk polycrystalline Pb(Zr$_{0.7}$Ti$_{0.3}$)O$_3$ to domain wall interactions with grain boundaries. Samples with grain sizes in the range of 3.9–10.4 µm were prepared and their microstructure, crystal structure, and dielectric/electromechanical properties were investigated. A decreasing grain size was accompanied by a reduction in large-signal electromechanical properties and an increase in small-signal dielectric permittivity. High-energy diffraction analysis revealed increasing microstrains upon decreasing the grain size, while piezoresponse force microscopy indicated an increased local coercive voltage near grain boundaries. The changes in properties were thus related to strained material volume close to the grain boundaries exhibiting reduced domain wall dynamics.

# 1. Introduction

Ferroelectric ceramics are widely applicable in electronic devices and are indispensable in actuators, high-dielectric-constant capacitors, piezoelectric sensors, and ultrasonic transducers [1-3]. Microstructural and crystallographic parameters, such as crystal structure [4], degree of crystallographic texture [5, 6], lattice distortion [7], and porosity [8] are commonly used to tailor their application-relevant electrical and electromechanical properties. Grain-size engineering is one of the most important tools and was previously used to tailor properties, such as permittivity [9], strain [10], fracture toughness [11], polarization reversal [12], electrical fatigue [13], high-power piezoelectric properties [14], and even phase transitions [15].

Dielectric and piezoelectric properties of ferroelectric ceramics have intrinsic and extrinsic contributions [16]. The former are related to the crystal lattice, while the latter refer to the movement of interfaces, i.e., domain walls or phase boundaries. For polycrystalline $Pb(Zr,Ti)O_3$ (PZT) ceramics with tetragonal and rhombohedral phase coexistence, extrinsic contributions are as high as 50–80% [17] of the overall dielectric permittivity and piezoelectric coefficient, while the value for single-phase compositions strongly depends on the crystal structure [18]. X-ray diffraction [19, 20], frequency-dependent dielectric measurements [21], and theoretical analysis [22] highlighted that the extrinsic contributions also display a strong grain size-dependence. For example, the peak in domain wall mobility for fine-grained $BaTiO_3$ (BT) ceramics at ~1 µm is the origin of their superior permittivity and piezoelectric properties [19, 21]. Grain size effects in PZT, however, seem to be fundamentally different [23] and less investigated. Limited understanding of the grain size effects is also related to the problem that the grain size decrease is accompanied by simultaneous changes in many other parameters; for example, changes in grain-to-grain-coupling [20], internal stresses [24], inhomogeneous local electric-field-distributions [25], and domain wall-grain boundary interactions [26]. In addition, the different grain sizes are typically achieved by varying the processing conditions, such as sintering temperature or dwell times, which can influence the chemical homogeneity and the type and concentration of charged point defects in the samples [27-29]. Therefore, all these parameters must be carefully considered in the interpretations of the effects of the grain size.

Recent studies [30-33] have revealed the importance of domain wall-grain boundary interactions. For example, piezoresponse force microscopy (PFM) was used to outline decreasing domain wall-related contributions in the vicinity of a 24° tilt grain boundary in a bicrystal thin film [30]. In the bulk, the domain wall-grain boundary interactions under an applied electric field were found to be dependent on the local mechanical strain energy. [31] Such spatial variations are related to the strong dependence of the grain boundary´s pinning potential on its type and misorientation angle. [32] In addition, some grain boundaries allow domain wall continuity [34], which might lead to enhanced or reduced domain wall mobility, related to correlated motion or pinning of domain walls, respectively. [33] While many fundamental studies investigated the interactions between single domain walls and grain boundaries [30-33, 35-37], this knowledge has not been transferred to polycrystalline materials, for which the amount of grain boundaries scales with the grain size. [38]

To elaborate on the domain wall-grain boundary interactions in polycrystalline ceramics, we investigated a series of PZT samples with a well-defined single-phase rhombohedral structure and grain sizes in the range of 3.9–10.4 µm. A previous study on a similar grain size range (2–10 µm) in polycrystalline $Pb(Zr_{0.415}Ti_{0.585})O_3$ reported on a continuous decrease of the remanent polarization and small-signal piezoelectric and dielectric properties [12], without identifying the microstructural origin of this decrease. In agreement, we find that decreasing the grain size comes along with a decrease in the switchable strain and polarization in our materials. By excluding other possible effects related to point defects and domain structure, we outline using a combination of X-ray diffraction (XRD) and PFM analyses that the changes in electromechanical and dielectric properties observed in the investigated materials with different grain sizes are related to the changes in microstrains, which impact the domain wall mobility in the vicinity of grain boundaries.

## 2. Experimental

### 2.1. Material synthesis and characterization

Rhombohedral Pb(Zr$_{0.7}$Ti$_{0.3}$)O$_3$ ceramic samples with different grain sizes were prepared by the mixed-oxide route. PbO (Sigma-Aldrich, MO, USA, 99.9%), ZrO$_2$ (Tosoh, Japan, 99.1%), and TiO$_2$ (Alfa Aesar, MA, USA, 99.8%) were mixed and homogenized in a planetary mill for 2 h at 200 min$^{-1}$ (PM400, Retsch GmbH, Germany) using an yttria stabilized zirconia (YSZ) vial, YSZ milling balls with diameter of 5 mm, and isopropanol as the dispersing medium. Afterwards the powders were dried at 95 °C and then calcined at 900 °C for 1 h in a closed alumina crucible using heating/cooling rates of 5 °C/min. This procedure was repeated two times. According to literature [29], calcination temperatures as high as 900 °C should ensure an efficient reaction of the raw materials to PZT. The calcined powders were attrition milled (Netzsch GmbH, Germany) in isopropanol at 800 min$^{-1}$ for 4 h using YSZ milling balls with the diameters of 3 mm. This was followed by drying the powders at 95 °C. Cylindrical pellets with a diameter of 8 mm were pressed uniaxially at 50 MPa, followed by cold isostatic pressing at 300 MPa. The samples were sintered in closed alumina crucibles with atmospheric powder with the same chemical composition at temperatures ranging from 1150 °C to 1250 °C, while the sintering time varied from 2 h to 8 h. The heating and cooling rates were 5 °C/min. Sintering temperatures above 1250 °C and prolonged sintering times were avoided to prevent considerable volatilization of PbO [28].

The density of the samples was measured using the Archimedes method. The theoretical density of 7.97 g/cm³ was used for calculating the relative densities [39]. The grain size was determined using Scanning Electron Microscopy (SEM, Jeol JSM-7600F, Japan). Therefore, the samples were ground, polished to 0.25 µm (DP-Paste M, Struers, Denmark), and chemically etched. For the grain size analysis, 250–500 grains were analyzed in each sample using the Image Tool Software [40]. The average grain size is expressed as the mean value of the Feret's diameter, while the range represents its standard deviation.

### 2.2. Synchrotron characterization

Diffraction experiments were carried out at the beamline ID15A [41] of the European Synchrotron Facility (ESRF) using a beam energy of 71.0 keV (λ=17.46 pm) in transmission geometry. Diffraction intensities were collected using the Pilatus CdTe 2M area detector (Dectris, Switzerland). The diffraction images from the detector were corrected for dead pixels, beam stop shadow, and image distortion. To calibrate X-ray energy and detector geometry, a reference measurement was carried out using a standard CeO$_2$ sample (National Institute of Standards and Technology, MA, USA).

Recorded patterns were fitted by the Rietveld method (details are provided in Supplementary Figure S1) using the program TOPAS 5. [42] Line profiles of the individual diffraction peaks were described by Pseudo-Voigt functions, accounting for specimen-related size and strain broadening, as well as the intrinsic broadening related to instrument resolution [43, 44]. The latter was estimated based on the refined pattern of $CeO_2$. For refinement of the PZT samples two additional parameters (one Gaussian, one Lorentzian) were used to model the broadening of peak profiles. Peak broadness, $\beta$, (determined as the Full-Width-Half-Maximum) expressing broadening due to crystallite size and microstrains was extracted after deconvolving the Pseudo-Voigt profile attributed to instrumental resolution. The low-angle values of $\beta$, extracted from Rietveld refinements (Tables S1-S4), reflect broadening due to combined size and strain. The $\tan(\theta)$ dependence of $\beta$, which is specific to microstrain, was extracted from single-peak fits and mapped in Williamson-Hall plots [45].

### 2.3. Electrical characterization

Before electrical characterization the samples were annealed at 500 °C for 1 h. A heating and cooling rate of 2 °C/min was used. The samples were allowed to relax and age for 25 days at room temperature after annealing before electrical characterization was carried out.

For impedance measurements platinum electrodes were sputtered on both surfaces of the disc-shaped samples after grinding. Impedance measurements were carried out using the Alpha-A impedance analyzer (Novocontrol Technologies, Germany) in the range from 600 °C to 200 °C (the cooling process was measured). An amplitude of 0.1 V and a measurement frequency range from 0.1 Hz to 3 MHz were used. The collected data was evaluated by the RelaxIS software (rhd instruments, Germany).

Temperature- and frequency-dependent relative permittivity was measured on annealed samples at frequencies from 1 kHz to 1 MHz using the impedance analyzer 4192A (Hewlett Packard Corporation, CA, USA). A voltage of 1 V, a heating rate of 2 °C/min, and a maximum temperature of 450 °C were used. The dielectric behavior was measured in two subsequent cycles to monitor the response of an aged (first cycle) and annealed (second cycle) sample state.

The strain and polarization measurements were performed with a modified Sawyer-Tower setup using a reference capacitor of 10 µF, equipped with an optical displacement sensor (Philtec Inc., MD, USA), and driven by a customized LabView program. A triangular wave signal with a frequency of 0.5 Hz and an amplitude of 3 kV/mm was applied with a voltage amplifier (Trek Model 20/20C, NY, USA). The first 2000 cycles were measured to relax the hysteresis [46] and access the true grain size-dependent large-signal electromechanical response.

## 2.4. PFM characterization

Prior to PFM investigations, sample surfaces were polished down to 0.25 µm with diamond paste, followed by 1 h of silica gel polishing in dilute KOH solution. The polished samples were heated to 500 °C for 1 h and then cooled to room temperature with a rate of 1 °C/min, to release internal stresses, which form due to sample preparation. The samples were investigated with an atomic force microscope (Molecular Force Probe 3D, CA, USA) equipped with a PFM module. A Ti/Ir-coated Si tip with a radius of curvature ~20 nm (Asyelec 01, AtomicForce F&E GmbH, Germany) was used for scanning. The out-of-plane PFM amplitude images were recorded in dual AC resonance tracking (DART) mode using AC sinusoidal signal with amplitude 7 V and a frequency of ~350 kHz. Domain wall density was determined by dividing the number of domain walls by the area of the PFM image. The local coercive voltages were measured using DART switching spectroscopy in off-loop mode. The applied signal was an increasing DC step signal, as reported elsewhere [47]. The sequence of increasing DC electric-field steps was 20 Hz with a maximum amplitude of 50 V. The frequency of the triangle envelope was 0.2 Hz, with an overlapping AC sinusoidal signal of 4 V and frequency of 350 kHz.

# 3. Results

## 3.1. Microstructure and crystal structure

The grain size of the analyzed PZT samples was controlled by varying the temperature and dwell time of the sintering process. Figure 1 displays the microstructures of the investigated PZT samples sintered at different conditions. The average grain size and the relative density of the samples are summarized in Table 1. The relative densities reached 96–98% of the theoretical density, irrespective of the sintering conditions. The average grain size increases from 3.9±1.8 µm to 10.4±4.0 µm, as the temperature and/or the dwell time increased. In addition, the distribution of grain sizes broadened, as evidenced by the increased standard deviation. For the subsequent discussion, the samples will be labelled as displayed in the first column of Table 1, whereby the number in the label denotes the average grain size (e.g., GS3.9 refers to the PZT sample with the average grain size of 3.9 µm).

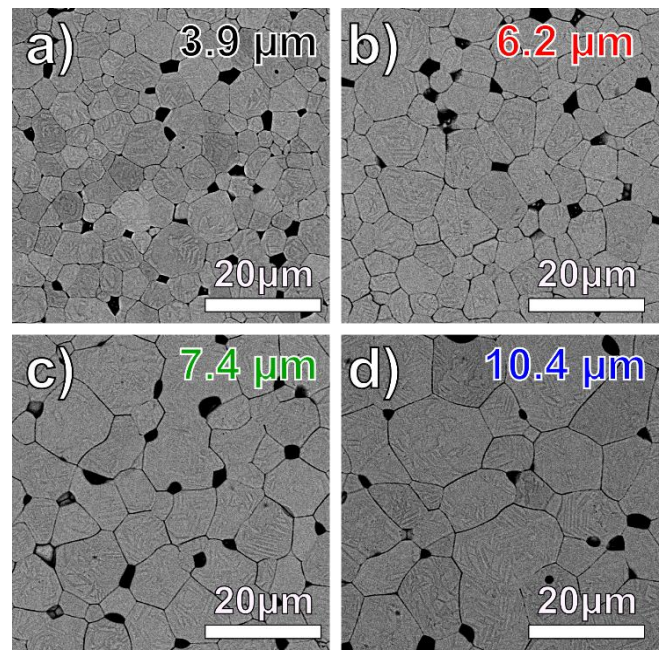

*Figure 1: Microstructures of etched PZT samples sintered at different conditions: a) 1100 °C, 2 h, b) 1150 °C, 2 h, c) 1150 °C, 8 h, and d) 1250 °C, 2h. The colored labels denote the average grain size.*

Table 1: Summary of crystallographic and microstructural parameters for samples with different grain sizes: average grain size ($g$), absolute ($\rho_{abs}$) and relative ($\rho_{rel}$) density, sintering temperature ($T$), sintering time ($t$), and microstrain ($\epsilon$) obtained from Williamson-Hall analysis. The abbreviation used for the samples in the text is displayed in the first column.

| Sample Label | $T$ (°C) | $t$ (h) | $\rho_{abs}$ (g/cm³) | $\rho_{rel}$ (%) | $g$ (μm) | $\epsilon, 10^{-3}$ |
|---|---|---|---|---|---|---|
| GS3.9 | 1100 | 2 | 7.80 | 97.8 | 3.9±1.8 | 1.21±0.09 |
| GS6.2 | 1150 | 2 | 7.71 | 96.7 | 6.2±2.4 | 1.14±0.09 |
| GS7.4 | 1150 | 8 | 7.62 | 95.6 | 7.4±3.1 | 1.13±0.09 |
| GS10.4 | 1250 | 2 | 7.69 | 96.4 | 10.4±4.0 | 1.11±0.09 |

The impact of grain size on the static ferroelectric domain structure in unpoled samples with the smallest (GS3.9) and the largest (GS10.4) grain sizes is presented in Figure 2 using PFM amplitude images. The red lines highlight the positions of the grain boundaries, which were obtained from the simultaneously measured topography contrast (not presented here). Both samples contain irregular and wedge-shaped domain walls, which was referred to as a the banded domain structure before. [48] According to microstructural analysis from previous studies, the irregular watermarks are 180° domain walls (green arrow in Fig. 2b), while the wedge-shaped features are likely non-180° domain walls (purple arrow in Fig. 2b) [49]. Note that a banded domain structure is typical for PZT with grain sizes larger than ~2 μm [50]. As expected [50], the domain structure of the GS3.9 sample is finer (1.94 domain walls/μm², Figure 2a) as compared to the structure of the GS10.4 sample (1.09 domain walls/μm², Figure 2b).

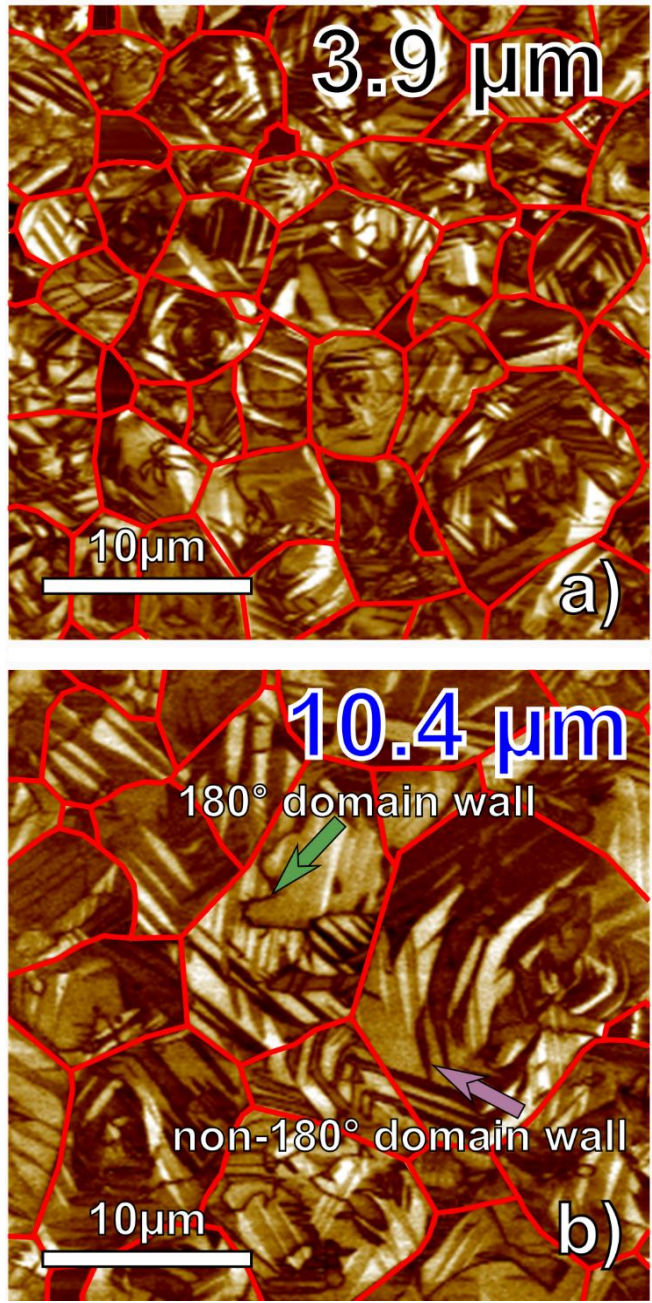

*Figure 2: PFM amplitude images of unpoled rhombohedral PZT samples with average grain sizes of a) 3.9 µm and b) 10.4 µm. The red lines indicate the position of the grain boundaries according to topography images taken simultaneously (not shown here). The banded domain structure [50], which contains 180° and non-180° domain walls is visualized by the respective green and purple arrow in b).*

Changes in grain size are known to influence the internal strains of polycrystalline ceramics [19, 20]. To this end, changes in the lattice parameters and microstrains of the PZT samples with different grain sizes were investigated using XRD. In order to avoid surface effects, high-energy X-ray synchrotron radiation was used in transmission geometry, providing average information from the bulk of the sample. Full XRD patterns are displayed in supplementary Figure S1, while the (111) and (200) diffraction peaks are highlighted in Figure 3 a. No peak shift or pronounced peak broadening can be observed at the first glance. This is related to the investigated range of grain size being well-above the submicron range, for which large changes were previously reported [19, 51]. To get a detailed view on the possible subtle changes, full-pattern Rietveld refinement of the diffractograms was carried out using the $R3c$ rhombohedral perovskite phase. The refined patterns and a complete set of structural parameters are summarized in Supplementary Figure S1 and Tables S1–S4. The differences in the calculated lattice distortions between the samples were negligible, indicating that the lattice strains remain almost unaffected in the investigated grain size range. Please note that significant changes in lattice strains were reported to occur for grain sizes below 4 µm in polycrystalline PZT ceramics. [23] Next, Williamson-Hall analysis [52] (Figure 3b) was used to quantify microstrains. The slope of the linear fit is proportional to the maximum microstrains in the sample. The differences in $\varepsilon$ between the samples, albeit close to the accuracy of limit of the method, point to a larger incidence of microstructural defects on the peak profile for smaller grain sizes, in particular the sample with the smallest grain size shows higher microstrain level than the rest of the samples with larger grain size. The inset in Figure 3b indicates that the mean values of microstrain increase with decreasing grain size. As displayed in Table 1, the microstrain ($\varepsilon, 10^{-3}$) of GS3.9 and GS10.4 is 1.21±0.09 and 1.11±0.09, respectively. Please note that, in agreement to our finding, a continuous increase of microstrains was previously reported for polycrystalline BT materials in the grain size range of 0.1–100 µm [53].

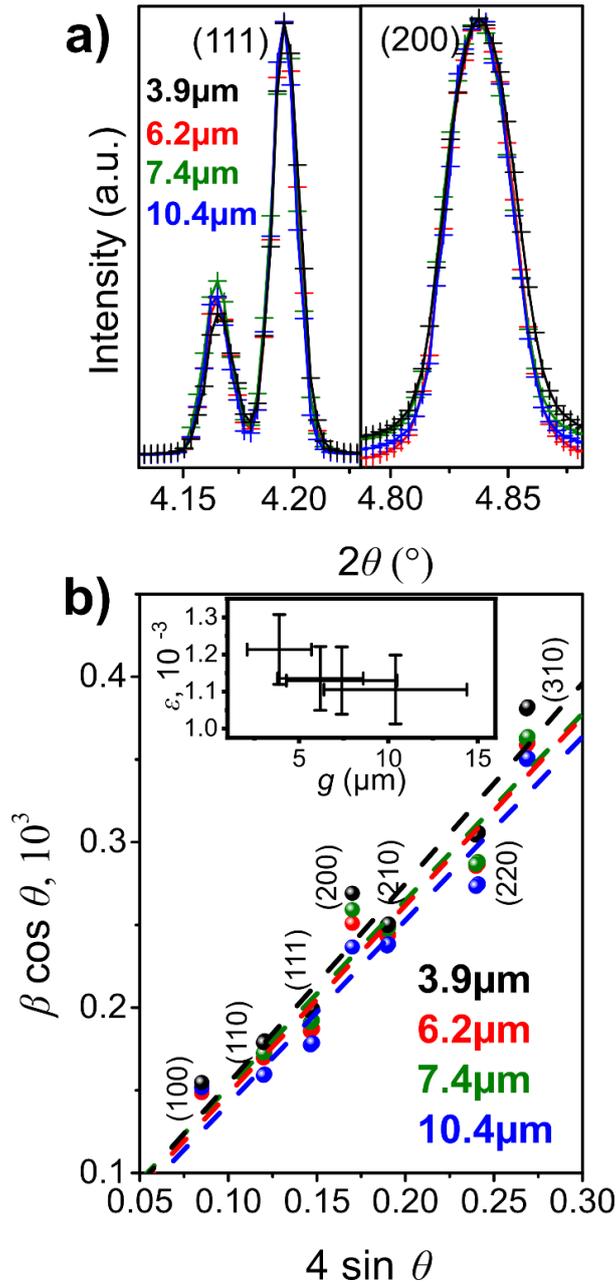

*Figure 3: High energy synchrotron XRD patterns of a) 111 and 200 Bragg reflections of polycrystalline rhombohedral PZT samples with different grain sizes. Full patterns and refined parameters can be found in supplementary Figure S1 and Tables S1-S4, respectively. The dashed line is a linear fit to the experimental data. The slope of the linear fit, which is proportional to the microstrain, ε, is displayed as a function of the grain size in b).*

The absence of secondary phases in XRD patterns for all grain sizes confirms that all samples are single phase. However, since the grain size was tailored by changing the sintering conditions, the samples were exposed to different temperatures for different dwell periods. Depending on these conditions, the PbO evaporation [28] leads to the formation of negatively-charged lead vacancies and positively-charged oxygen vacancies. By forming defect complexes, these defects are known to affect the electrical and electromechanical properties [54, 55], thus the estimation of their concentration is crucial.

Due to experimental limitations, it is difficult to measure the extent of PbO evaporation directly. However, changes of ionic conductivity measured by impedance spectroscopy can be used to estimate PbO evaporation indirectly [56]. Temperature-dependent impedance measurements on the PZT samples with different grain sizes are reported and discussed in supplementary Figure S2. The data reveal negligible differences in the conductivity between the samples with different grain size, indicating that the concentration of lead and oxygen vacancies in the investigated samples is comparable, within the experimental limitation. The observed changes in electromechanical and dielectric properties (discussed in Section 3.2) can thus be solely related to microstructural effects due to a change in grain size and not due to defect-related phenomena.

3.2. Electrical and electromechanical properties

In order to reveal if aging impacts the small signal dielectric response, two subsequent heating/cooling measurement cycles were carried out starting with an aged sample. Note that no differences were found between both cycles for all grain sizes (not shown here), indicating that the state of the defect complexes does not impact the dielectric response of the samples. Figure 4 displays dielectric permittivity and loss data of the second measurement cycle, representing an annealed sample. The permittivity value at the phase transition temperature drops by 40% with decreasing the grain size, as displayed in Figure 4a and Table 2. This is in good agreement with previous studies on PZT [23]. The distribution of the permittivity values around the $T_C$ can be quantified by comparing the peak´s Full-Width-at-Half-Maximum (FWHM) value. The FWHM increases from 24 °C to 38 °C upon decreasing the grain size from 10.4 µm to 3.9 µm (see Table 2), which indicates a broader distribution of permittivity values around the $T_C$. Further information on the distribution of permittivity values at lower temperatures can be obtained by scaling the permittivity axis logarithmically (Figure 4b). Interestingly, the permittivity decrease at $T_C$ comes along with a permittivity increase at lower temperatures (indicated by the arrows in Figure 4b). The room temperature permittivity was found to increase by 50% upon decreasing the grain size from 3.9 µm to 10.4 µm. Albeit smaller, this effect is fundamentally similar to BT materials [9]. In BT these scaling effects were related to a peak in domain-wall related contribution with changing the grain size. [19, 21]

As displayed in Figure 4c, the $T_C$ was determined as the intersection of the slopes from the temperature variation of the $1/\epsilon$ curves, following Ref. [57]. The $T_C$ does not vary with grain size in the investigated range. In agreement with this, no significant variation of the $T_C$ with grain size was reported for sizes between 0.6 µm and 10.9 µm for lanthanum doped PZT [58]. The ratio of the $1/\epsilon$ slopes below and above $T_C$ is obtained from Figure 4c and displayed in Table 2. The ratio was previously used to quantify the order of the phase transition. Values of -8 and -2 were suggested for first and second order transitions, respectively [59]. The values reported in Table 2 indicate that the phase transition is second order [60] and that the grain size does not impact the phase transition order.

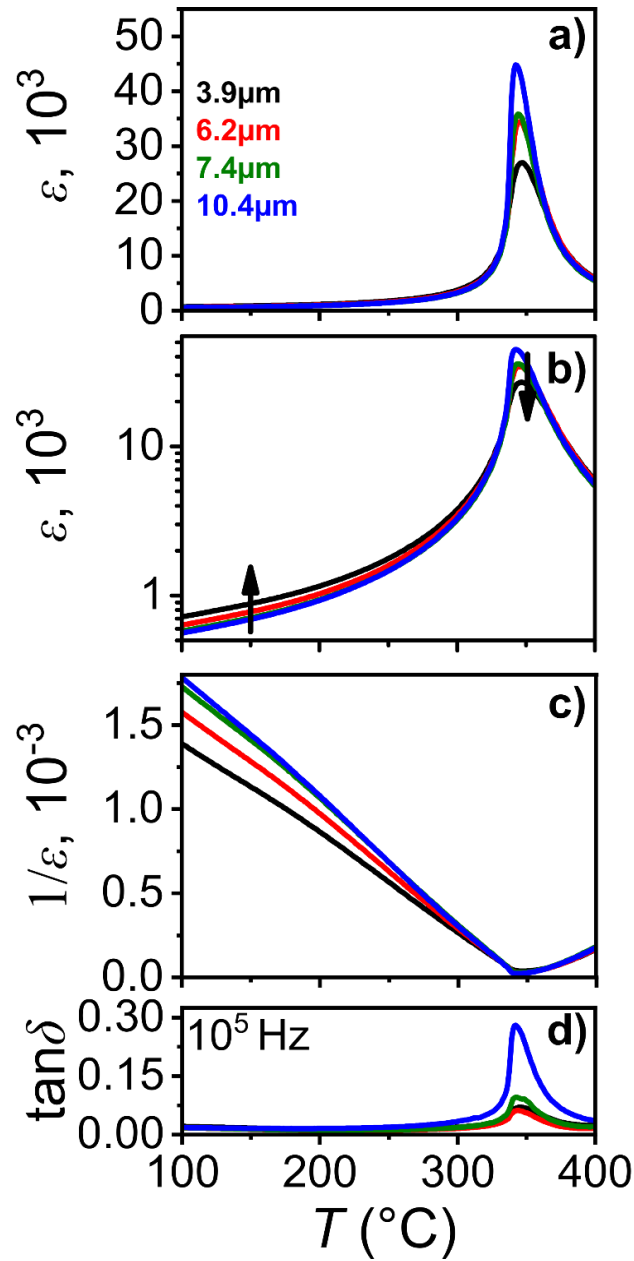

Figure 4: Temperature-dependent a,b) permittivity, c) reciprocal permittivity, and d) loss factor of annealed Pb(Zr$_{0.7}$Ti$_{0.3}$)O$_3$ ceramic samples (2$^{nd}$ measurement cycle) with different grain sizes (f=10$^5$ Hz). Please note that the permittivity in b) is scaled logarithmically. Influences of grain size on the permittivity at the Curie temperature and at lower temperatures are visualized by arrows.

Large-signal polarization and strain hysteresis loops of polycrystalline Pb(Zr$_{0.7}$Ti$_{0.3}$)O$_3$ with different grain sizes were measured during 2000 cycles, starting from an aged state. Loops after different number of cycles are displayed in supplementary Figure S3. The 1$^{st}$ cycle polarization loops of all samples are pinched and no negative strain can be observed in the strain hysteresis. This indicates pinning effects from point defects, likely oxygen-vacancy-related defect complexes [54], which align with the polarization direction during aging and reduce the DW mobility. The polarization loop pinching continuously disappears with AC electric field cycling (supplementary Figure S4). However, details of this behavior will not be further investigated, as the scope of this work is the investigation of the grain size influence. All the measurements discussed below were thus conducted on de-aged samples. To this end, comparing the response after 2000 cycles (Figure 5, Table 2) reveals a bipolar strain, $S_{bip}^{2000}$, of 0.12% for the GS3.9 and 0.16% for the GS10.4 sample (30% relative increase). Also, the negative strain, $S_{neg}^{2000}$, increases with increasing the grain size, i.e., from 0.04% for GS3.9 to 0.07% for GS10.4 (75% relative increase). This comes along with an increase in the remanent polarization, $P_r^{2000}$, from 22.8 µC/cm² to 25.0 µC/cm² for GS3.9 and GS10.4, respectively (10% relative increase). Note that small differences in $P_r^{2000}$ were found for the samples GS6.2, GS7.4, and GS10.4 (values vary within a maximum relative difference of ~4%). The switchable polarization, $\Delta P^{2000}$ (determined at ±3 kV/mm) behaves similar.

*Table 2: Small-signal (Curie temperature, T$_c$, maximum permittivity at T$_c$ and room temperature (RT), $\epsilon_{33}$, FWHM at T$_c$, ratio of $1/\epsilon_{33}$ slopes below and above T$_c$) and large-signal (bipolar $S_{bip}^{2000}$, and negative $S_{neg}^{2000}$ strain after 2000 cycles) properties of PZT with different grain sizes.*

| Sample Label | T$_c$ (°C) | $\epsilon_{33} \cdot 10^3$ (RT, 10$^5$ Hz) | $\epsilon_{33} \cdot 10^3$ (T$_c$, 10$^5$ Hz) | FWHM (°C) | $\dfrac{\left(\dfrac{d\epsilon^{-1}}{dT}\right)_{T<T_c}}{\left(\dfrac{d\epsilon^{-1}}{dT}\right)_{T>T_c}}$ | $S_{bip}^{2000}$ (%) | $S_{neg}^{2000}$ (%) | $E_c^{2000}$ (kV/mm) | $P_r^{2000}$ (µC/cm²) | $\Delta P^{2000}$ (µC/cm²) |
|---|---|---|---|---|---|---|---|---|---|---|
| GS3.9 | 351 | 0.6 | 27.0 | 38 | -1.3 | 0.12 | 0.04 | 0.67 | 22.8 | 61.5 |
| GS6.2 | 351 | 0.5 | 34.3 | 31 | -1.6 | 0.14 | 0.05 | 0.74 | 25.4 | 64.3 |
| GS7.4 | 351 | 0.5 | 35.7 | 28 | -1.6 | 0.15 | 0.06 | 0.76 | 26.0 | 64.9 |
| GS10.4 | 349 | 0.4 | 44.5 | 24 | -1.7 | 0.16 | 0.07 | 0.72 | 25.0 | 64.4 |

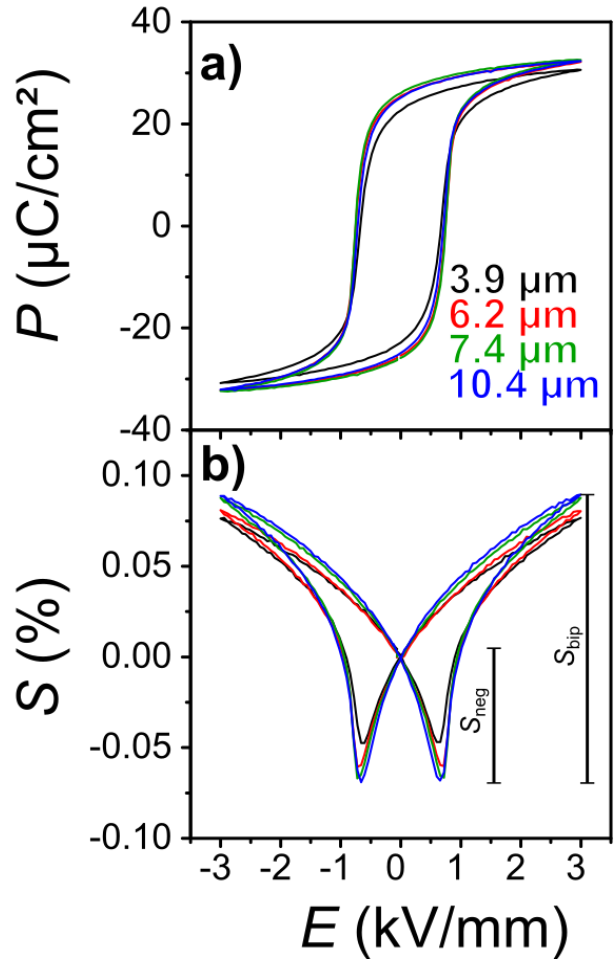

*Figure 5: a) Polarization and b) strain hysteresis loops for de-aged Pb(Zr$_{0.7}$Ti$_{0.3}$)O$_3$ ceramic samples with different grain sizes after 2000 cycles. The definition of the bipolar (S$_{bip}$) and negative (S$_{neg}$) strain is given in b).*

### 3.3. Local polarization switching measurements

Measurements of the macroscopic electrical properties reveal a clear grain size dependence (Table 2), while synchrotron diffraction analysis indicates that the microstrain in GS3.9 is 10% higher in comparison to GS10.4 counterparts. In order to localize changes in domain wall mobility in the material, complementary measurements using GS10.4 as an exemplary sample were carried out by PFM. To this end, the local coercive voltages were mapped spatially resolved with a 20-nm tip resolution. The location of measurement points is displayed as black dots in Figure 6a. Coercive voltages were obtained by averaging positive and negative voltages from local bipolar hysteresis loops from the third cycle. Hysteresis loops are displayed in the supplementary (Figure S5). Local coercive voltages are plotted as a contour plot in 6b. Note that 10 grains were measured across the microstructure and only representative examples are displayed here, while others can be found in supplementary Figure S6. Areas containing high domain wall densities are observed in the vicinity of grain boundaries (indicated by arrows in Figure 6a). Moreover, these areas with high domain wall density are found to be directly related to the amplitude of the local coercive voltage, whereby higher values of coercive voltages were observed in the areas with higher domain wall density (Figure 7 b). In agreement to previous reports investigating local switching, volumes near grain boundaries containing high domain wall density [30, 31] were found to be characterized by high coercive voltage needed for local domain switching. [61]

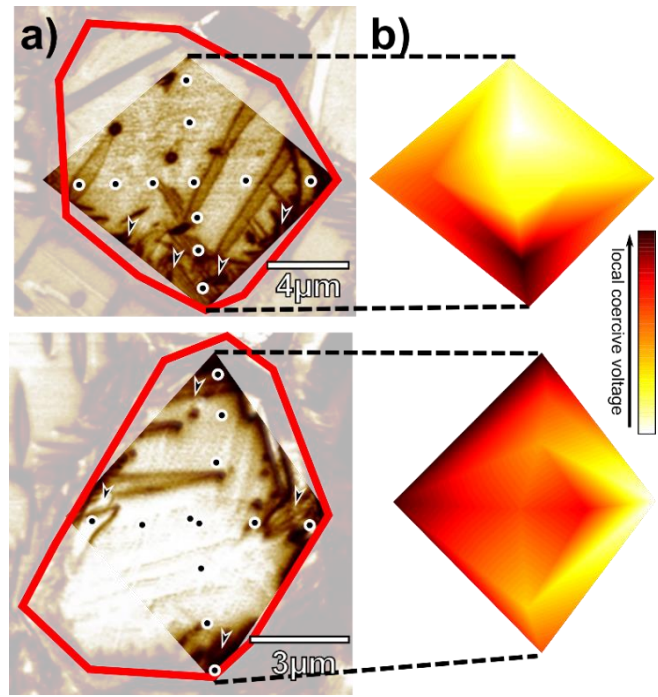

*Figure 6: a) PFM image of the domain structure of two representative grains of a sample with the average grain size of 10.4 µm. The grain boundary is displayed as a red line, while the measurement points are marked by black dots. The arrows indicate the position of areas with a high density of domain walls. b) Average local coercive voltages obtained from the third measurement cycle (exemplary PFM phase-field responses are displayed in supplementary (Figure S5)). Red and yellow colors in b) indicate areas of high and low coercive voltage, respectively (see color bar). Additional maps of the local coercive voltage are provided in supplementary (Figure S6).*

## 4. Discussion

### 4.1. Point Defects

Processing conditions are known to decisively influence the functional properties of PZT ceramics [27-29]. Pinched polarization hysteresis loops were previously reported for PZT materials and were related to the presence of oxygen-vacancy-containing defect complexes [54]. This behavior is also observed in this work, irrespective of the sample´s grain size (supplementary Figure S3). In the aged state the defect complexes align along the polarization direction and impose a hindering and restoring force during polarization reversal, which is on the macroscopic level observed as pinched polarization loops. Therefore, as evident from the pinched bipolar polarization loops and the absence of the negative strain (supplementary Figure S3), defects strongly impede the movement of domain walls [62]. Continuous bipolar cycling results in de-aging (hysteresis relaxation), whereby the initially ordered defect complexes probably rearrange into a more disordered state, as indicated by first principle studies [63]. In such a state, the defect complexes do not have the same pinning effect and the macroscopic polarization loops display typical ferroelectric behavior, i.e., the loops de-pinch and open. In the present work, the characteristic onset cycle for de-aging was found to be relatively independent of the grain size (supplementary Figure S4), indicating similar defect-mediated DW pinning effects and possibly similar defect concentration [64, 65]. To further elaborate the latter, impedance spectroscopy measurements were carried out. As evident from supplementary Figure S2, the activation energy for the bulk DC conductivity is equivalent for all the samples, indicating comparable defect states in the investigated samples. Note that an increased defect concentration, which could be expected for the samples with the larger grain size due to the prolonged sintering time or higher temperature, would result in a decrease of the electromechanical properties, which was clearly not observed (Table 2). Nevertheless, to avoid the influence from defects, the samples were de-aged using 2000 large-field cycles. This measure, along with the results from the impedance spectroscopy, justifies to relate the observed changes in further experiments to the differences in grain sizes.

### 4.2. Residual strains

The average grain size of the investigated samples ranges from 3.9 µm to 10.4 µm. Materials within this grain size range are expected to exhibit a banded domain structure in the virgin state [50], which was confirmed by the PFM analysis (Figure 2). While 180° and non-180° domain walls can be observed in all samples, the domain wall density increases by 80% and the domain size decreases with decreasing the grain size from 10.4 µm to 3.9 µm. This trend is typically related to an interplay between elastic and electrical energy, as well as the domain wall energy [50].

The impact of grain size on the lattice strain and microstrains was investigated using high-energy X-ray radiation, which allows the sampling of the samples´ bulk and thus avoiding surface effects. Decreasing the size from 10.4 µm to 3.9 µm results in a relative increase of the microstrains by 10%, while at the same time the lattice distortion remains almost unchanged (Supplementary Table S1–S4). In agreement to this finding, polycrystalline BT displays a similar increase of microstrains on a grain size range of 0.1–100 µm [53]. Strained volumes in polycrystalline ceramic materials are known to occur in the vicinity of interfaces [66], such as domain walls [67] and grain boundaries [68]. While the current experimental diffraction data (Figure 3) does not allow to locate the strained volumes in the investigated samples, previous temperature-dependent XRD measurements on polycrystalline BT (temperature range 25–600 °C) indicated that strained volumes are most likely located in the vicinity of grain boundaries [53]. Decreasing the grain size results in an increased number of grain boundaries and thus a higher volume fraction of material in the grain boundary vicinity. Microstrains in the vicinity of grain boundaries were intensively studied in structural ceramics and were related to the anisotropy in the thermal expansion coefficient and a random orientation of the grains. For example, in polycrystalline $Al_2O_3$ the strain concentration close to the grain boundary exceeds the value in the center of the grain by a factor of 4 [69]. In ferroic materials, an additional source is the formation of the spontaneous strain at the phase transition temperature [70]. In a recent study on polycrystalline $(Ba,Ca)(Zr,Ti)O_3$ [31], it was found that absolute elastic strains are significantly enhanced in the vicinity of the grain boundary (4–5 µm) and decrease towards the center of the 33 µm large grain.

4.3.    Domain-wall-grain boundary interactions

Strained volumes in the vicinity of grain boundaries are likely to exhibit a high domain wall density. For example, a spatial correlation between volumes of high elastic mechanical energy and high domain wall density was recently established. [31] These volumes are expected to exhibit different domain wall dynamics – a similar interplay has been previously suggested by theory [61]. In the present work, this interaction was studied using local coercive voltage mapping (Figure 6 and supplementary Figure S6). It was demonstrated that the local coercive voltage is significantly enhanced in grain boundary regions exhibiting a high domain wall density. This finding is in good agreement to experiments on polycrystalline PZT thin films. In fact, local hysteresis loops measured in the grain boundary vicinity were reported to exhibit a strong imprint [26, 36], while the corresponding coercive voltage was found to be larger compared to the center of the grain [35]. In agreement, electric field-dependent transmission electron microscopy outlined that the reorientation of domains in the vicinity of the grain boundary remains incomplete even under high electric fields [71]. The influenced volume is difficult to estimate, as it depends on multiple parameters, including material composition, boundary orientation angle, and temperature [32]. Also note that some grain boundaries also allow domain wall continuity that can impact the influenced volume [33]. In rhombohedral materials, it has been highlighted that for high symmetry grain-grain misorientations, such as $\sum 3$ or $\sum 7$, domain continuity through the boundary is very likely regardless of the grain boundary plane, while for random grain-grain misorientations, domain continuity is highly dependent on the grain boundary plane and thus the ability of the material to compensate ferroelastic strain mismatch and electrical charge [72]. In an exemplarily study, a 24° tilt grain boundary was recently investigated by Marincel et al. in a rhombohedral $Pb(Zr_{0.52}Ti_{0.48})O_3$ bicrystal thin film. They found that that domain-wall motion was impacted in a range of $450 \pm 30$ nm in the vicinity of a 24° tilt grain boundary [73].

It should be noted that PFM is a surface sensitive method. This is particularly important for bulk ferroelectric/ferroelastic materials, as their properties strongly depend on the mechanical boundary conditions, which vary from plane strain in the bulk to plane stress at the surface [74]. This problem can be overcome by the recent advances in three-dimensional X-ray diffraction microscopy [31], which enables spatially-resolved mapping of domains with sub-100 nm resolution within the bulk. Reduced domain wall dynamics in the vicinity of the grain boundary at electric fields exceeding ten times the coercive field were observed using this technique in polycrystalline $(Ba,Ca)(Zr,Ti)O_3$. Nevertheless, the qualitative agreement between both measurement techniques confirms that PFM results can be used as a guideline for investigating polycrystalline ceramic materials.

As discussed above, the reduced grain size was demonstrated to increase the volume fraction of grain boundary regions, where domain wall dynamics are impacted. These local phenomena were found to have a pronounced influence on the macroscopic dielectric (Figure 4, Table 2) and electromechanical properties (Figure 5, Table 2). A summary is provided in Figure 7. The dielectric permittivity at the phase transition temperature of the sample GS3.9 is 40% lower compared to the GS10.4 sample. At the same time, the room temperature permittivity is enhanced by 50%. A similar, however more pronounced, trend is frequently reported in polycrystalline BT materials [9]. While the small-signal dielectric and piezoelectric properties usually peak at about 1 µm in BT ceramics with decreasing grain size, the large-signal properties continuously decrease. [75, 76] In agreement to BT, the bipolar and negative strain of the investigated PZT materials decrease by 25% and 40%, respectively, while the remanent polarization decreases by 10% with decreasing the grain size (Table 2, Figure 7). Note that a continuous decrease of electromechanical properties was previously also reported for PZT materials with grain sizes in the range of 2.0–9.8 µm, without discussing the origin of the changes on the microstructural level. [12] Also studies using surface XRD methods demonstrated that in PZT materials the decrease in large-signal properties is accompanied by a decreased contribution of non-180° domain wall motion [20, 77].

The results presented in this work suggest that the changes in electrical and electromechanical large-signal properties in polycrystalline PZT ceramics are mainly related to a change in the strain state in the grain boundary vicinity (Figure 3c), impacting the domain wall density and locally limiting the domain wall movement (Figure 6). This can explain the large signal properties, such as reduction in bipolar strain and remanent polarization, with decreasing grain size (see Figure 7a and b). However, the question as to the origin of the increasing room-temperature permittivity upon reducing the grain size remains open (see Figure 7c, black curve). Frequency-dependent permittivity measurements [21] and in situ XRD data [19] indicate that in BT this trend is related to increased domain wall contributions. A similar mechanism may be thus inferred also in PZT, i.e., the increased domain wall density (Figure 2 and Figure 6) may be responsible for the increased domain wall contribution and increased room temperature permittivity at small grain sizes, e.g., via domain wall vibrations under small applied fields. On the other hand, the domain switching process, involving domain nucleation and growth [62], might be more strongly affected by the grain boundaries. As a result, large signal parameters decrease with decreasing grain size.

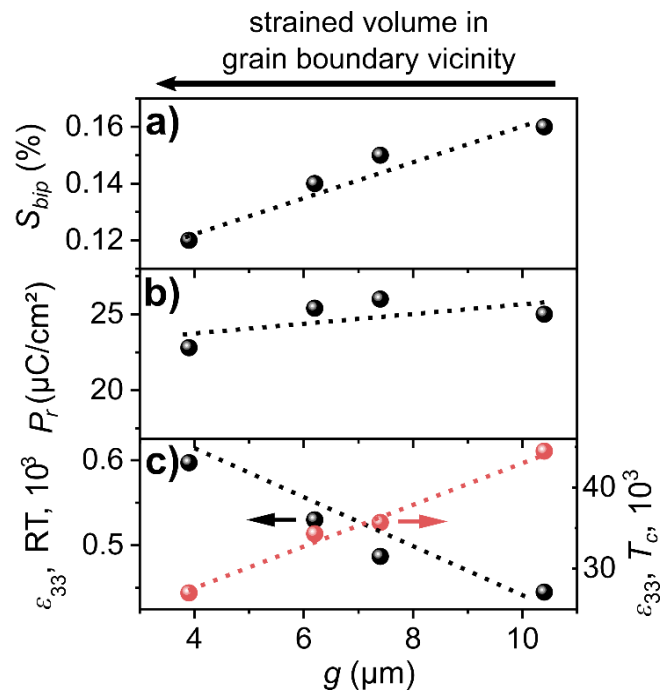

Figure 7: Impact of grain size on large-signal a) bipolar strain, $S_{bip}$, b) remanent polarization, $P_r$, and small-signal c) dielectric permittivity, $\epsilon_{33}$, at room temperature and at $T_c$ for polycrystalline Pb(Zr$_{0.7}$Ti$_{0.3}$)O$_3$. Experimental data are summarized in Table 2. The dashed

*lines display a guideline to the eye. The increasing fraction of strained volume in the grain boundary vicinity is indicated by the arrow.*

## 5. Conclusions

Despite the impact of grain size on the macroscopic electromechanical and dielectric properties being well documented in polycrystalline Pb(Zr,Ti)$O_3$ ceramics, discussions and throughout explanations on the local origin of the observed effects remain an open question. [10, 23] Recent studies [30, 31, 33], however, brought the grain boundary-domain wall interaction into focus, without discussing its impact of macroscopic properties in polycrystalline ceramic materials. Here, we present a synergetic approach combining macroscopic electromechanical and dielectric measurements, with bulk synchrotron diffraction and highly localized scanning probe techniques.

The results are in qualitative agreement to the scaling effects observed in polycrystalline BaTi$O_3$ [9, 75]. Large-signal electromechanical properties continuously decrease with decreasing grain size (negative strain by 40%, bipolar strain by 25%, remanent polarization by 10%), while the small-signal permittivity at the Curie temperature decreases by 40%, accompanied by a 50% increase at room temperature. Rietveld refinement of the high-energy x-ray diffraction data revealed that the microstrains in the investigated materials increase by 10% with decreasing grain size. The differently-strained volumes are located in the vicinity of the grain boundaries and locally impact the movement of domain walls, as directly confirmed by PFM measurements of the local coercive voltages. The domain wall-grain boundary interactions are thus identified as the main reason for the decrease in large-signal functional properties, such as the remanent polarization and the bipolar switching strain with decreasing grain size.

**Acknowledgement**

This work was supported by the Deutsche Forschungsgemeinschaft (DFG) Grant No. KO 5100/1-1. We are grateful to the ESRF for providing beamtime at ID15A. J.S. acknowledges the support of the Feodor Lynen Research Fellowship Program of the Alexander von Humboldt. Parts of the work were also supported by the DAAD through funds from the Bundesministerium für Bildung und Forschung (BMBF) Grant No. 57402439 and the Slovenian Research Agency through the bilateral project PR-08298 (contract number BI-DE/18-19-007). We are grateful to S. Drnovšek for preparing the samples.

# Supplementary material

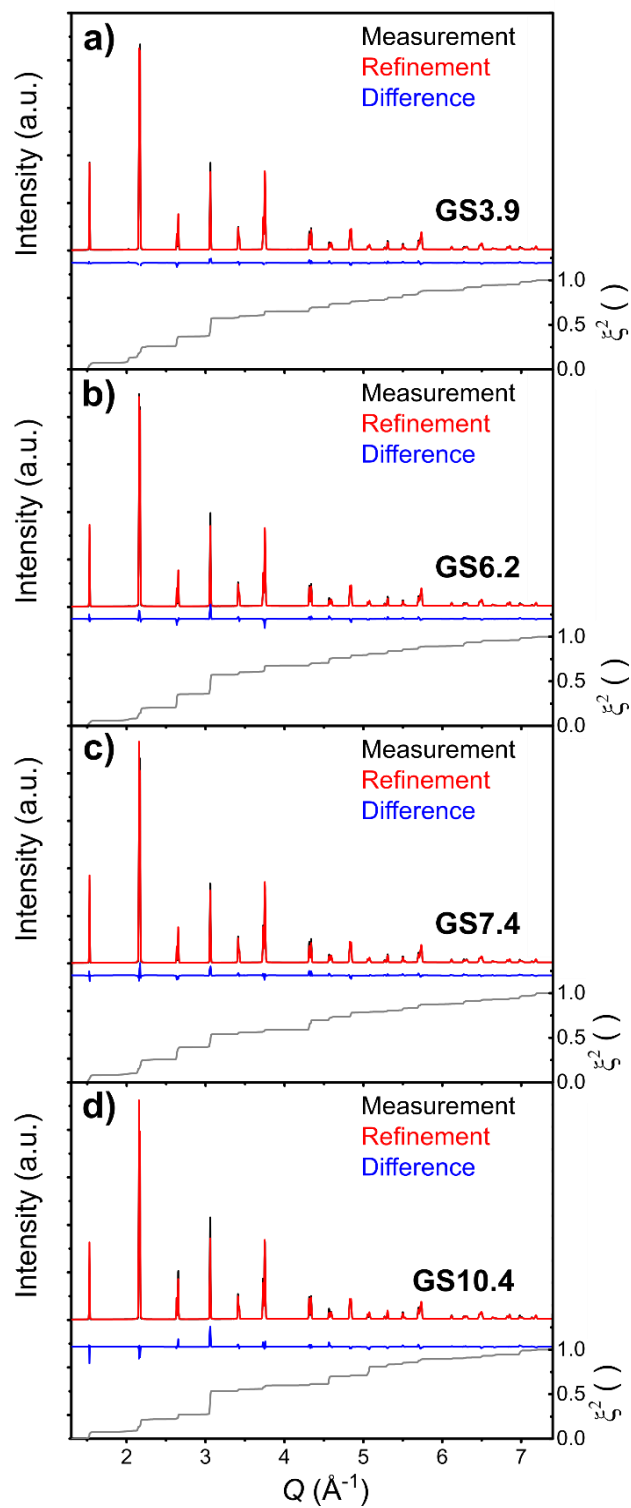

*Figure S1: Raw XRD data obtained from synchrotron diffraction (black line), refined data (red line), and difference between raw and refinement data (blue line) displayed for polycrystalline samples with different grain sizes: a) GS3.9, b) GS6.2, c) GS7.4, and d) GS10.4. The $\xi^2$ plot quantifies the quality of the fit for each Q value. The refinements of each*

PZT sample and $CeO_2$ included a background profile (5 parameters) and one isotropic thermal parameter for each atom type. Each PZT phase also had variable lattice parameters and variable z coordinate of the perovskite A and B sites. Incorrect fit of certain peak intensities (e.g. the $024_r$ (=$200_c$)) is probably related to preferred orientation of crystallites in the sample, which could not be spun due to its geometry, and mainly affects the thermal factors.

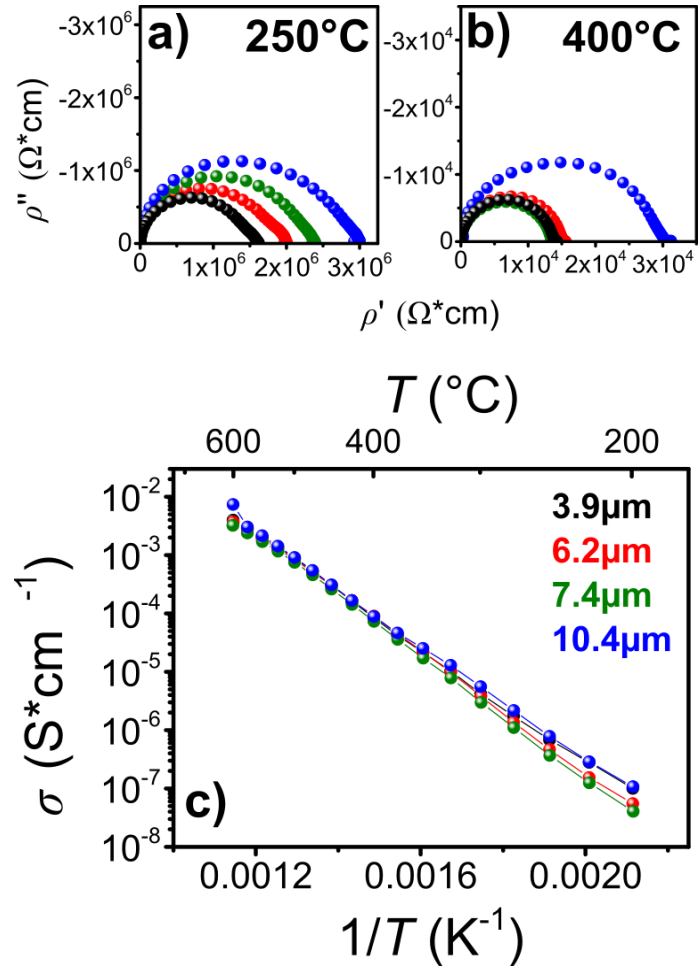

Figure S2: Complex impedance plots (imaginary part, $\rho''$ against real part, $\rho'$, of impedance) of $Pb(Zr_{0.7}Ti_{0.3})O_3$ ceramic samples with different grain sizes at various measurement temperatures: a) 250 °C and b) 400 °C. One dominant semicircle could be identified for all samples, indicating the bulk process. In addition, an overlapping smaller low-frequency process was observed at lower temperatures, which can most likely be attributed to an electrode response. Arrhenius-type plots for the bulk DC conductivity are displayed in c). The bulk conductivity of undoped and donor doped PZT is expected to be mixed ionic-electronic in nature [1-3]. Our results in c) show that the activation energy for the bulk conduction process is 0.97 eV for all samples, irrespective of the grain size. Since much

*higher activation energies of 2.36 eV were previously reported for oxygen vacancy migration in similar PZT compositions [3], the minor differences in conductivity between the samples at lower temperatures (T<400 °C) are most likely due to small changes in electronic contribution. On the other hand, the high-temperature (T>400 °C) conductivity, to which oxygen transport would contribute the most, remains unchanged. Previous research where oxygen transport is assumed as the dominant conduction process in PZT illustrated a high activation energy at low temperatures and a lower activation energy at high temperatures hinting towards the formation of defect couples. [4] This is clearly not observed for the investigated samples in c), which indicates that the concentration of lead and oxygen vacancies in the investigated samples is comparable, within the experimental limitation. The observed changes in electromechanical and dielectric properties (discussed in Section 3.2) can thus be solely related to microstructural effects due to a change in grain size.*

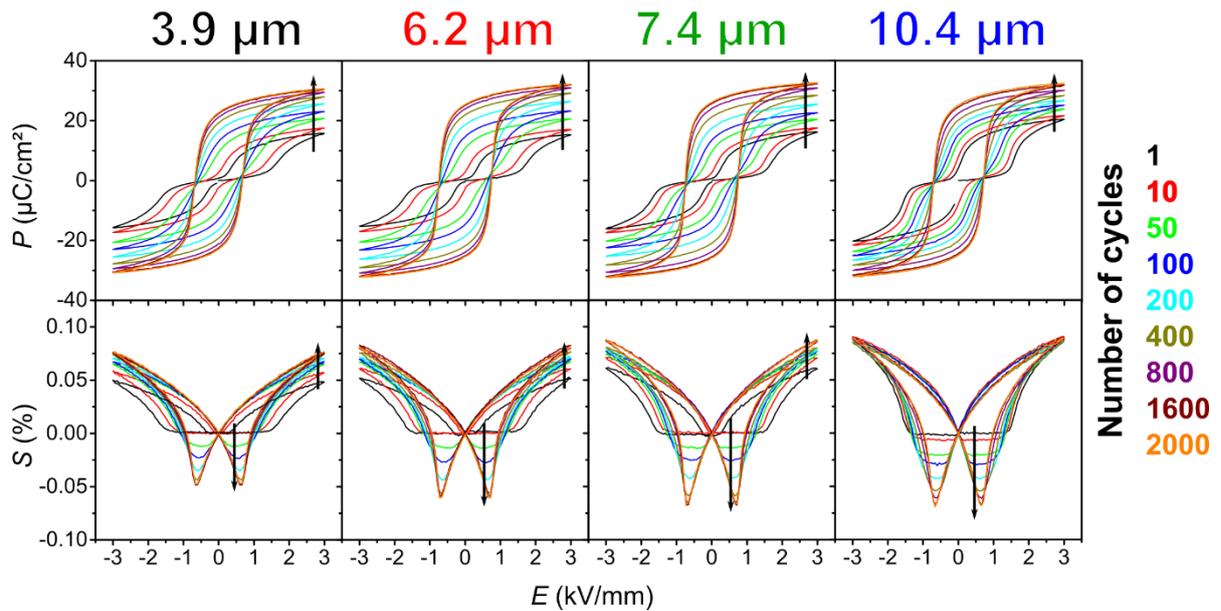

*Figure S3: Relaxation of the polarization and strain hysteresis loops for $Pb(Zr_{0.7}Ti_{0.3})O_3$ ceramic samples with different grain sizes. Cycles 1, 10, 50, 100, 200, 400, 800, 1600, and 2000 are displayed. The arrows indicate the influence of the increasing number of cycles on the saturation polarization and the negative strain.*

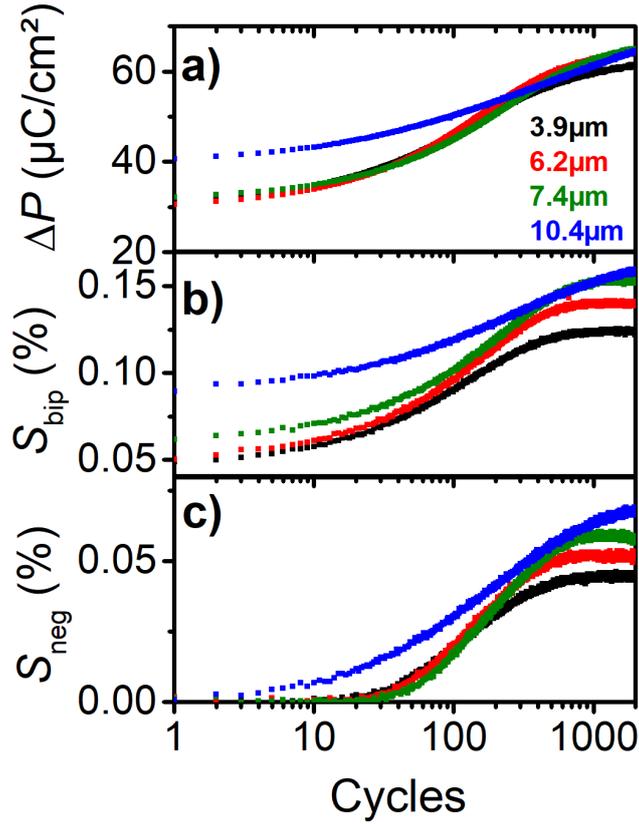

*Figure S4: Switching-cycle dependence of the a) switchable polarization (quantified from polarization hysteresis loops at 3kV/mm), b) bipolar strain, $S_{bip}$, and c) negative strain, $S_{neg}$. Bipolar and negative strain are determined according to the definition given in Figure 5 b.*

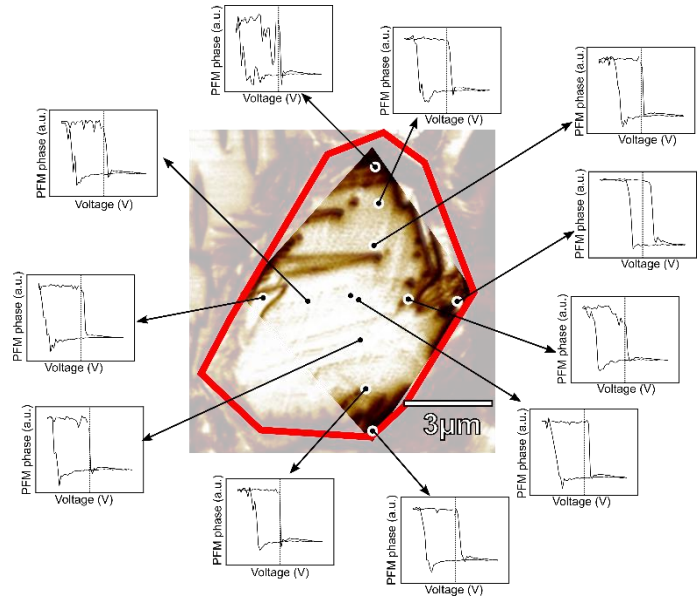

*Figure S5: PFM phase response as a function of the applied voltage for different positions within the grain. Please note that all PFM amplitude responses have the same x-scale. The third cycle is displayed. The spatial resolution of the coercive voltage (displayed in Figure 6*

*and Figure S6) was obtained by averaging the positive and negative coercive voltage for each position.*

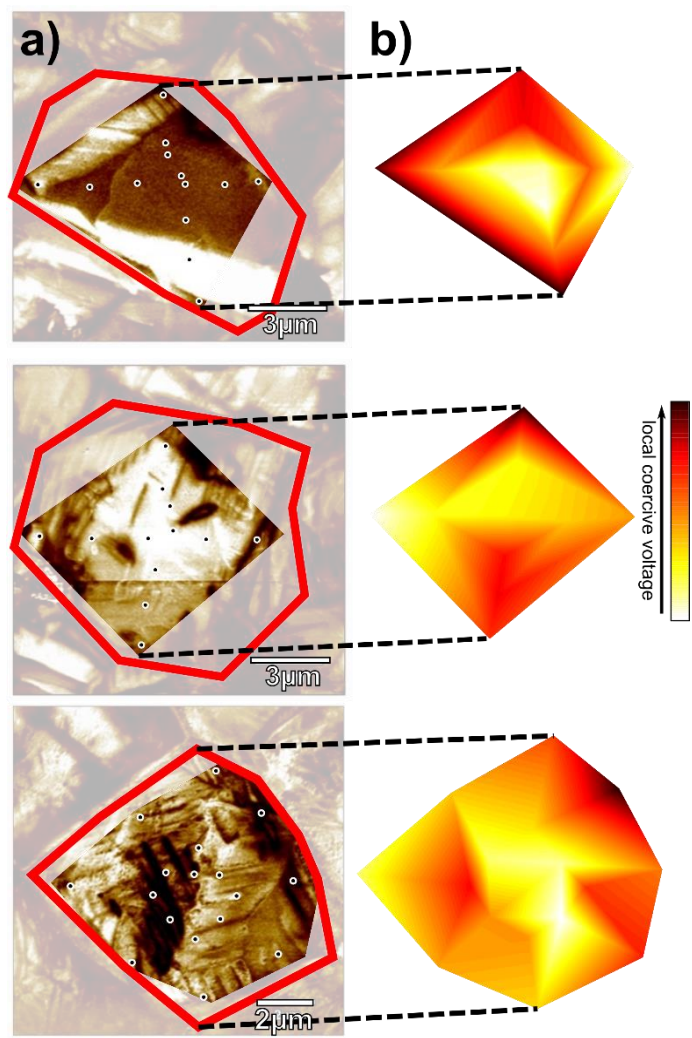

*Figure S6: Further examples for the interplay between domain structure and local coercive field (extension of Figure 6).*

Table S1: Structural information from Rietveld refinement of GS3.9

| Space group *R3c*, *a* = 5.7944(1) Å, *c* = 14.3062(3) Å, $R_{wp}$=9.74% | | | | | |
|---|---|---|---|---|---|
| | *x* | *y* | *z* | occ [Zr/Ti] | $B_{iso}$ (Å$^2$) |
| Pb | 0 | 0 | 0.281(1) | 1 | 2.19(3) |
| Zr/Ti | 0 | 0 | 0.049(1) | 0.7/0.3 | 0.32(6) |
| O | 0.1719 | 0.3706 | 0.0796 | 1 | 4.8(4) |
| $c_{pc}/a_{pc}$ | 1.00796(3) | | | | |
| Pseudo-Voigt FWHM | 0.0085(1) | | | | |

Table S3: Structural information from Rietveld refinement of GS6.2

| Space group *R3c*, *a* = 5.7946(1) Å, *c* = 14.3082(3) Å, $R_{wp}$=11.0% | | | | | |
|---|---|---|---|---|---|
| | *x* | *y* | *z* | occ [Zr/Ti] | $B_{iso}$ (Å$^2$) |
| Pb | 0 | 0 | 0.271(1) | 1 | 2.05(4) |
| Zr/Ti | 0 | 0 | 0.037(1) | 0.7/0.3 | 0.33(7) |
| O | 0.1719 | 0.3706 | 0.0796 | 1 | 5.1(4) |
| $c_{pc}/a_{pc}$ | 1.00806(3) | | | | |
| Pseudo-Voigt FWHM | 0.0078(1) | | | | |

Table S2: Structural information from Rietveld refinement of GS7.4

| Space group *R3c*, *a* = 5.7955(1) Å, *c* = 14.3103(3) Å, $R_{wp}$=10.2% | | | | | |
|---|---|---|---|---|---|
| | *x* | *y* | *z* | occ [Zr/Ti] | $B_{iso}$ (Å$^2$) |
| Pb | 0 | 0 | 0.284(1) | 1 | 2.09(3) |
| Zr/Ti | 0 | 0 | 0.052(1) | 0.7/0.3 | 0.37(6) |
| O | 0.1719 | 0.3706 | 0.0796 | 1 | 6.0(5) |
| $c_{pc}/a_{pc}$ | 1.00805(3) | | | | |
| Pseudo-Voigt FWHM | 0.0080(1) | | | | |

Table S4: Structural information from Rietveld refinement of GS10.4

| Space group *R3c*, *a* = 5.7947(1) Å, *c* = 14.3091(3) Å, $R_{wp}$=12.4% | | | | | |
|---|---|---|---|---|---|
| | *x* | *y* | *z* | occ [Zr/Ti] | $B_{iso}$ (Å$^2$) |
| Pb | 0 | 0 | 0.271(1) | 1 | 1.81(4) |
| Zr/Ti | 0 | 0 | 0.038(1) | 0.7/0.3 | 0.28(8) |
| O | 0.1719 | 0.3706 | 0.0796 | 1 | 0.8(3) |
| $c_{pc}/a_{pc}$ | 1.00810(3) | | | | |
| Pseudo-Voigt FWHM | 0.0071(1) | | | | |